\documentclass[dvips,prd,preprint,amssymb,nofootinbib,
letterpaper]{revtex4-1}
\usepackage{amssymb}
\usepackage[dvips]{graphicx}
\usepackage{setspace}
\begin{document}
\preprint{}

\title{Dark matter annihilation through a lepton-specific Higgs boson}

\author{Heather E.\ Logan}
\email{logan@physics.carleton.ca}

\affiliation{Ottawa-Carleton Institute for Physics,
Carleton University, Ottawa K1S 5B6 Canada}

\date{November 1, 2010}

\begin{abstract}
It was recently argued by Hooper and Goodenough~\cite{Hooper:2010mq} 
that the excess gamma ray emission from within 1--2$^{\circ}$ of the
galactic center can be well-described by annihilation of $\sim$8~GeV
dark matter particles into tau pairs.  I show that such a dark matter
signal can be obtained naturally in the lepton-specific
two-Higgs-doublet model extended by a stable singlet scalar dark
matter candidate.  The favored parameter region prefers a light Higgs
state (below 200~GeV) with enhanced couplings to leptons and sizable
invisible branching fraction.  Part of the favored region leads to
invisible decays of both of the CP-even neutral Higgs states.
\end{abstract}

\maketitle

\section{Introduction}

Measurements of cosmic large-scale structure, galaxy and
galactic-cluster dynamics, and gravitational lensing overwhelmingly
support the existence of a non-baryonic, electrically neutral
(``dark'') component of the matter density of the universe.  The
observed dark matter density is consistent with its thermal production
in the early universe if its mass and annihilation cross section to
Standard Model (SM) particles are around the electroweak scale.  The
search for this weak-scale dark matter by means other than
gravitational makes up a sizable fraction of the global experimental
particle physics program.

Recently, Hooper and Goodenough~\cite{Hooper:2010mq} presented an
independent analysis of gamma ray data from within 10$^{\circ}$ of the
galactic center collected over the past two years by the Fermi Gamma
Ray Space Telescope.  They showed that the spatial distribution and
energy spectrum of gamma rays between about 1.25$^{\circ}$ and
10$^{\circ}$ of the galactic center are well described by a fit to a
galactic disk profile and a spherically symmetric galactic bulge, with
gamma rays originating from decays of neutral pions (produced in
hadronic cosmic ray interactions) and inverse Compton scattering
together with identified point sources.

This fit fails when extrapolated inward below 1.25$^{\circ}$.  Hooper
and Goodenough identified a new component, highly peaked at the galactic
center but not point-like.\footnote{After removing this new
  component, the spectrum of the point source at the galactic center
  was extracted and is consistent with the extrapolation of the power
  law spectrum observed by ground-based gamma ray telescopes.}  They
argued that an astrophysical origin for this new component is
implausible, and that instead the energy spectrum is very well fitted
by annihilation of dark matter particles of mass 7.3--9.2~GeV into
$\tau$ pairs.  Up to 10--20\% of annihilations can be into $b \bar b$
or $c \bar c$ (which yield softer gammas) without degrading the fit.
They further fit the spatial distribution of the new component
assuming annihilation (i.e., assuming a rate proportional to the square of
the number density) and find an inner density profile of $\rho \sim
r^{-1.34 \pm 0.04}$.  Assuming that this profile continues out to the
distance of the sun and normalizing to the local dark matter density,
the fitted rate implies a dark matter annihilation cross section of
$\langle \sigma v_{\rm rel} \rangle \simeq 3.3 \times
10^{-27}$~cm$^3$/s.  Assuming instead that the profile is more peaked
at the center of the galaxy due to attraction of the dark matter by
the very high baryonic matter density there, and that the density
profile softens to $\rho \sim r^{-1}$ outside the galactic core, they
find an annihilation cross section of $1.5 \times 10^{-26}$~cm$^3$/s.
These values are in rough agreement with the early universe cross section
required to obtain a thermal relic with the measured abundance,
$\langle \sigma v_{\rm rel} \rangle \simeq 3 \times
10^{-26}$~cm$^3$/s.  The moderate discrepancy could be accounted for if
the annihilation cross section is velocity dependent (such as for
$P$-wave annihilations) or if there are additional annihilation modes,
e.g., to electrons, muons, or neutrinos, which do not contribute
significantly to the gamma ray signal.

A particularly simple model for dark matter is a stable,
gauge-singlet, real scalar particle that annihilates through couplings
to the Higgs sector~\cite{Silveira:1985rk}.  Such a model was
described recently in the current context in
Ref.~\cite{Barger:2010mc}, in which a (complex) singlet scalar dark
matter particle with mass of 10--30~GeV annihilates through
$s$-channel SM Higgs exchange to SM fermions.  Because ratios of the
SM Higgs couplings to fermions are fixed by the fermion masses,
annihilations into $b \bar b$ final states dominate for dark matter
particle masses from just above the $b$ threshold up to about 65~GeV
where annihilations to off-shell $W$ bosons begin to turn on.  In
particular, a dominant annihilation mode to $\tau$ pairs can only be
achieved if the couplings of the relevant exchanged particles can be
made to favor the leptonic final state.

With this motivation I consider the annihilation of a gauge-singlet
scalar dark matter candidate through interactions with the
``lepton-specific'' two-Higgs-doublet model (2HDM-L), in which one
doublet generates masses for the SM quarks while the other generates
masses for the leptons.  This model has previously been considered in
the context of dark matter annihilations in Ref.~\cite{Goh:2009wg},
which studied a variety of dark matter extensions of the 2HDM-L with
several-hundred-GeV dark matter particles annihilating or decaying
into 2HDM-L states in order to explain the PAMELA and ATIC high-energy
positron and electron excesses.  Here I focus on annihilation of
$\sim$8~GeV dark matter particles through $s$-channel exchange of the
two CP-even Higgs bosons of the 2HDM-L.  I show that an appropriate
annihilation cross section into $\tau$ pairs and sufficient
suppression of hadronic final states can be achieved through
appropriate choice of the model parameters.  The dark matter
annihilation cross section fitted in Ref.~\cite{Hooper:2010mq} favors
a mass for the ``leptonic'' Higgs below 200--300 GeV.  I also show
that part of the preferred parameter region leads to large invisible
decay branching fractions for both of the CP-even neutral Higgs
bosons.  Even though the dark matter particle receives contributions
to its mass through couplings to the two Higgs doublets, its light
mass can be obtained without too much fine tuning through appropriate
choices for the couplings and vacuum expectation values of the two
doublets.  

This paper is organized as follows.  In Sec.~\ref{sec:model} I present
the model and summarize the relevant couplings.  In
Sec.~\ref{sec:dmann} I compute the dark matter annihilation cross
sections to $\tau$ pairs and quark pairs and determine the parameters
necessary to agree with the fitted annihilation cross section of
Ref.~\cite{Hooper:2010mq}.  In Sec.~\ref{sec:naturalness} I consider
the question of naturalness of the small mass of the dark matter
candidate.  In Sec.~\ref{sec:invisihiggs} I compute the invisible
Higgs decay branching fractions and comment on discovery prospects in
various parts of parameter space.  In Sec.~\ref{sec:direct} I comment
on the spin-independent dark matter direct detection cross section in
this scenario.  Conclusions are summarized in
Sec.~\ref{sec:conclusions}.

\section{The model}
\label{sec:model}

The lepton-specific two-Higgs-doublet model contains two complex
SU(2)-doublet scalar fields, $\Phi_q$ and $\Phi_{\ell}$, where
$\Phi_q$ generates the masses of both up- and down-type quarks while
$\Phi_{\ell}$ generates the masses of the charged leptons.  This
Yukawa coupling structure was first introduced in
Refs.~\cite{Barnett:1983mm,Barnett:1984zy,Grossman:1994jb} and the
couplings, decays, and phenomenology of the resulting Higgs states
have been studied in
Refs.~\cite{AkeroydLEP,Akeroyd:1998ui,Hmodels,Goh:2009wg,Aoki:2009ha,
Su:2009fz,Logan:2009uf,Cao:2009as}.  This Yukawa structure was also
used in the neutrino mass model of Ref.~\cite{Aoki:2008av}.

The two doublets can be written as,
\begin{equation}
   \Phi_i = \left( \begin{array}{c} \phi_i^+ \\
     \frac{1}{\sqrt{2}} \left( \phi_i^{0,r} + v_i + i \phi_i^{0,i} \right) 
   \end{array} \right), \qquad\qquad 
   i = q, \ell.
\end{equation}
The vacuum expectation values are constrained by $v_q^2 + v_{\ell}^2 =
v_{\rm SM}^2 = (246~{\rm GeV})^2$, with their ratio parameterized as
\begin{equation}
  \tan\beta \equiv \frac{v_q}{v_{\ell}}.
\end{equation}
The desired form of the Yukawa Lagrangian is enforced by imposing a
$Z_2$ symmetry, broken only softly in the Higgs potential, under which
$\Phi_{\ell}$ and the right-handed leptons transform as $\Phi_{\ell}
\to -\Phi_{\ell}$ and $e_{Ri} \to -e_{Ri}$,
with all other fields unaffected.  The Yukawa Lagrangian is then,
\begin{equation}
  \mathcal{L}_{\rm Yuk} = -\sum_{i,j=1}^3 \left[ 
   y_{ij}^{u} \overline{u}_{Ri} \widetilde{\Phi}_q^{\dagger} Q_{Lj}
 + y_{ij}^{d} \overline{d}_{Ri} \Phi_{q}^{\dagger} Q_{Lj}
 + y_{ij}^{\ell} \overline{e}_{Ri} \Phi_{\ell}^{\dagger} L_{Lj} \right]
 + {\rm h.c.},
\label{eq:Lyuk}
\end{equation}
where $i,j$ are generation indices, $y^{u,d,\ell}_{ij}$ are the Yukawa 
coupling matrices, the left-handed quark and lepton doublets are given by
$Q_{Li} = (u_{Li}, d_{Li})^T$ and $L_{Li} = (\nu_{Li}, e_{Li})^T$,
and the conjugate Higgs doublet is,
\begin{equation}
  \widetilde{\Phi}_q \equiv i \sigma_2 \Phi_q^*
  = \left( \begin{array}{c}
  \frac{1}{\sqrt{2}} \left( \phi_q^{0,r} + v_q - i \phi_q^{0,i} \right) \\
  -\phi_q^- 
  \end{array} \right).
\end{equation}
The scalar potential for $\Phi_{q,\ell}$ was given, e.g., in Eq.~(2) of
Ref.~\cite{Su:2009fz}. 

To this model I add a gauge-singlet real scalar field $S$, which is
odd under a second, unbroken global $Z_2$ symmetry.  This second $Z_2$
ensures that $S$ is stable.  The Higgs potential acquires the
following new terms:
\begin{equation}
  V_S = \frac{1}{2} m_3^2 S^2 
  + \frac{\lambda_q}{2} S^2 \Phi_q^{\dagger} \Phi_q
  + \frac{\lambda_{\ell}}{2} S^2 \Phi_{\ell}^{\dagger} \Phi_{\ell} 
  + \lambda_S S^4.
  \label{eq:VS}
\end{equation}
Note that a term $\sim S^2 \Phi_q^{\dagger} \Phi_{\ell}$ is forbidden
by the requirement that the $Z_2$ responsible for the form of the
Yukawa Lagrangian is only softly broken.

After electroweak symmetry breaking, the couplings $\lambda_{q, \ell}$
lead to three-point couplings of two $S$ particles to the CP-even
neutral states in $\Phi_{q,\ell}$.  We parameterize the CP-even
neutral mass eigenstates in the usual way as
\begin{eqnarray}
  h^0 &=& -\sin\alpha \, \phi_{\ell}^{0,r} 
  + \cos\alpha \, \phi_q^{0,r}, \nonumber \\
  H^0 &=& \cos\alpha \, \phi_{\ell}^{0,r} + \sin\alpha \, \phi_q^{0,r}.
\end{eqnarray}
The Feynman rules for the couplings of these states to quarks, leptons, 
$W$ and $Z$ bosons, and $S$ pairs are then,
\begin{eqnarray}
  h^0 q \bar q &:& -i \frac{m_q}{v_{\rm SM}} \frac{\cos\alpha}{\sin\beta},
  \qquad \qquad
  h^0 \ell \bar \ell \, : \, i \frac{m_{\ell}}{v_{\rm SM}} 
  \frac{\sin\alpha}{\cos\beta}, 
  \qquad \qquad
  h^0 V_{\mu} V_{\nu} \, : \, 
  2 i \frac{M_V^2}{v_{\rm SM}} \sin(\beta - \alpha) 
  g_{\mu\nu},
  \nonumber \\
  h^0 SS &:& -i v_{\rm SM} ( \lambda_q \sin\beta \cos\alpha 
  - \lambda_{\ell} \cos\beta \sin\alpha),
  \nonumber \\
  H^0 q \bar q &:& -i \frac{m_q}{v_{\rm SM}} \frac{\sin\alpha}{\sin\beta},
  \qquad \qquad
  H^0 \ell \bar \ell \, : \, -i \frac{m_{\ell}}{v_{\rm SM}}
  \frac{\cos\alpha}{\cos\beta},
  \qquad \qquad
  H^0 V_{\mu} V_{\nu} \, : \, 
  2 i \frac{M_V^2}{v_{\rm SM}} \cos(\beta - \alpha) 
  g_{\mu\nu},
  \nonumber \\
  H^0 SS &:& -i v_{\rm SM} 
  (\lambda_q \sin\beta \sin\alpha + \lambda_{\ell} \cos\beta \cos\alpha).
  \label{eq:hcoups}
\end{eqnarray}
In what follows I will consider as an example the limit $\cos(\beta -
\alpha) = 0$, i.e., $\cos\alpha \to \sin\beta$ and $\sin\alpha \to -
\cos\beta$.  In this limit the couplings become,
\begin{eqnarray}
  h^0 q \bar q &:& -i \frac{m_q}{v_{\rm SM}},
  \qquad \qquad
  h^0 \ell \bar \ell \, : \, -i \frac{m_{\ell}}{v_{\rm SM}}, 
  \qquad \qquad
  h^0 V_{\mu} V_{\nu} \, : \, 
  2 i \frac{M_V^2}{v_{\rm SM}} 
  g_{\mu\nu},
  \nonumber \\
  h^0 SS &:& -i v_{\rm SM} ( \lambda_q \sin^2\beta 
  + \lambda_{\ell} \cos^2\beta ),
  \nonumber \\
  H^0 q \bar q &:& i \frac{m_q}{v_{\rm SM}} \cot\beta,
  \qquad \qquad
  H^0 \ell \bar \ell \, : \, -i \frac{m_{\ell}}{v_{\rm SM}}
  \tan\beta,
  \qquad \qquad
  H^0 V_{\mu} V_{\nu} \, : \,   0,
  \nonumber \\
  H^0 SS &:& -i v_{\rm SM} 
  (\lambda_{\ell} - \lambda_q) \sin\beta \cos\beta.
  \label{eq:hcoupscba}
\end{eqnarray}

\section{Dark matter annihilation signature}
\label{sec:dmann}

In this model, the annihilation of two $S$ particles proceeds via
$s$-channel exchange of $h^0$ and $H^0$.  Neglecting the kinetic
energies of the nonrelativistic initial-state dark matter particles,
the annihilation cross section into fermions $f \bar f$ is given by,
\begin{equation}
  \sigma v_{\rm rel} = \frac{N_c m_f^2}{4\pi} C_f^2 
  \left[1 - \frac{4 m_f^2}{s}\right]^{3/2},
  \label{eq:sigmav}
\end{equation}
where $v_{\rm rel}$ is the relative velocity of the two $S$ particles,
$N_c$ is the number of colors of fermion species $f$, $s$ is the square of
the center-of-mass energy (equal to $4 M_S^2$ in the nonrelativistic limit),
and $C_f$ is the relevant product of coupling and propagator factors given 
for leptons and quarks by,
\begin{eqnarray}
  C_{\ell} &=& 
  \frac{(\lambda_q \tan\beta \cos\alpha - \lambda_{\ell} \sin\alpha) 
    \sin\alpha}{s - M_h^2}
  - \frac{(\lambda_q \tan\beta \sin\alpha + \lambda_{\ell} \cos\alpha)
    \cos\alpha}{s - M_H^2}
  \nonumber \\
  C_q &=& 
  -\frac{(\lambda_q \cos\alpha - \lambda_{\ell} \cot\beta \sin\alpha)
    \cos\alpha}{s - M_h^2}
  - \frac{(\lambda_q \sin\alpha + \lambda_{\ell} \cot\beta \cos\alpha)
    \sin\alpha}{s - M_H^2}.
\label{eq:Cs}
\end{eqnarray}
Here $M_h$ and $M_H$ are the masses of $h^0$ and $H^0$, respectively.
Note that averaging over the dark matter particle velocity distribution
has no effect on this cross section because the cross section
is velocity-independent in the low-velocity limit.

Taking $M_S = 8$~GeV, the cross section for $SS \to \tau\tau$ yields the
constraint
\begin{equation}
  C_{\ell}^2 = \left(\frac{1}{128~{\rm GeV}}\right)^4 
  \left[ \frac{\sigma v_{\rm rel}}{10^{-26}~{\rm cm^3/s}}\right].
  \label{eq:Clvalue}
\end{equation}
Recall that the fit in Ref.~\cite{Hooper:2010mq} found an annihilation
cross section to $\tau\tau$ in the range $3.3 \times 10^{-27}$ to
$1.5 \times 10^{-26}$~cm$^3$/s.

The fit in Ref.~\cite{Hooper:2010mq} requires that $b \bar b$ and $c \bar
c$ final states make up no more than 10--20\% of annihilations, i.e.,
\begin{equation}
  \frac{\sigma v_{\rm rel}(SS \to b \bar b + c \bar c)}
       {\sigma v_{\rm rel}(SS \to \tau\tau)} \lesssim 0.2
\end{equation}
If the coupling factors $C_{\ell}$ and $C_q$ were equal, the ratio of
annihilation cross sections given above would be equal to the ratio of
the corresponding decay widths of a SM-like Higgs boson of mass $2M_S$.
Taking $M_S = 8$~GeV (the relevant SM Higgs branching ratios do not
vary much over the favored dark matter mass range of 7.3--9.2~GeV), I
find using the public code {\tt HDECAY} version 3.53~\cite{HDECAY},
\begin{equation}
  \frac{\Gamma(H_{\rm SM} \to b \bar b + c \bar c)}
  {\Gamma(H_{\rm SM} \to \tau \tau)} = 13.3.
\end{equation}
This yields an upper bound on the ratio of coupling factors,
\begin{equation}
  \frac{C_q^2}{C_{\ell}^2} \lesssim \frac{0.2}{13.3} = 0.015, \qquad
  {\rm or} \qquad \left| \frac{C_q}{C_{\ell}} \right| \lesssim 0.12,
  \label{eq:Cqlimit}
\end{equation}
implying that $C_q$ must be suppressed by about an order of magnitude
relative to $C_{\ell}$.  To characterize the parameter region in which
the excess gamma ray signature can be accommodated, I examine
$C_{\ell}$ and $C_q$ in two limits.

First, consider the case when $\phi_{\ell}^{0,r}$ and $\phi_q^{0,r}$
are mass eigenstates.\footnote{This basis can be automatically chosen
  if $h^0$ and $H^0$ are degenerate.}  Then the coupling factors
become,
\begin{equation}
  C_{\ell} = \frac{\lambda_{\ell}}{M_{\phi_{\ell}^{0,r}}^2 - s}, 
  \qquad \qquad
  C_q = \frac{\lambda_q}{M_{\phi_q^{0,r}}^2 - s}.
\end{equation}
In particular, $SS \to \ell \bar \ell$ proceeds only through
$\phi_{\ell}^{0,r}$ exchange with $SS$ coupling $\lambda_{\ell}$, and
$SS \to q \bar q$ proceeds only through $\phi_q^{0,r}$ exchange
with $SS$ coupling $\lambda_q$.  Taking $\lambda_q$ small is enough to
suppress the $q \bar q$ final states.

The required value of $C_{\ell}$ from Eq.~(\ref{eq:Clvalue}) yields a
relation between $\lambda_{\ell}$ and the $\phi_{\ell}^{0,r}$ mass;
neglecting $s = (16~{\rm GeV})^2$ relative to $M_{\phi_{\ell}^{0,r}}^2$,
I find,
\begin{equation}
  \lambda_{\ell} \simeq 
  \left[ \frac{M_{\phi_{\ell}^{0,r}}}{128~{\rm GeV}} \right]^2
  \left[ \frac{\sigma v_{\rm rel}}{10^{-26}~{\rm cm^3/s}} \right]^{1/2}.
\label{eq:lambdaleigen}
\end{equation}
In particular, the required dark matter annihilation cross section is
naturally obtained with $\lambda_{\ell} \sim 1$ and the
$\phi_{\ell}^{0,r}$ mass around 130~GeV.  For other masses, the
required coupling $\lambda_{\ell}$ scales proportional to the square
of the $\phi_{\ell}^{0,r}$ mass.  The upper bound on hadronic final
states from Eq.~(\ref{eq:Cqlimit}) leads to the constraint,
\begin{equation}
  \lambda_q \lesssim 0.12
  \frac{M_{\phi_q^{0,r}}^2}{M_{\phi_{\ell}^{0,r}}^2} \lambda_{\ell}.
\end{equation}
Taking $\lambda_{\ell}$ and $M_{\phi_{\ell}^{0,r}}$ as in
Eq.~(\ref{eq:lambdaleigen}), the required suppression can be achieved
either by making $\lambda_q$ small or by making $\phi_q^{0,r}$ heavy; 
$\lambda_q \sim 1$ requires $M_{\phi_q^{0,r}} \gtrsim 370$~GeV.

Second, consider the case of $\cos(\beta - \alpha) \to 0$.  In this
limit, the couplings of $h^0$ to SM particles are the same as those of
the SM Higgs and the $H^0 WW$ and $H^0 ZZ$ couplings are zero [see
Eq.~(\ref{eq:hcoupscba})].  The coupling and propagator factors become,
\begin{eqnarray}
  C_{\ell} &=& \frac{\lambda_q \sin^2\beta + \lambda_{\ell} \cos^2\beta}
  {M_h^2 - s}
  + \frac{ (\lambda_{\ell} -\lambda_q) \sin^2\beta}
  {M_H^2 - s}
  \nonumber \\
  C_q &=& \frac{\lambda_q \sin^2\beta + \lambda_{\ell} \cos^2\beta}
  {M_h^2 - s}
  - \frac{(\lambda_{\ell} - \lambda_q) \cos^2\beta}
  {M_H^2 - s}.
  \label{eq:Ccba}
\end{eqnarray}
In this case, taking $\lambda_q$ small is not enough to suppress the 
$q \bar q$ final states.  One also needs the quantity
\begin{equation}
  \lambda_{\ell} \cos^2\beta \left[ \frac{1}{M_h^2 - s} - \frac{1}{M_H^2 - s}
    \right]
\end{equation}
to be small, without making $\lambda_{\ell}$ small.  This can be
achieved by taking $M_H \to M_h$ or by taking $\tan\beta$ large
($\tan\beta \gtrsim 3 \sqrt{\lambda_{\ell}}$ is sufficient).  Large
$\tan\beta$ corresponds to making $H^0$ ``lepton-friendly,'' as can be
seen from the couplings to fermions given in Eq.~(\ref{eq:hcoupscba}).
In the large $\tan\beta$ limit I obtain,
\begin{equation}
  C_{\ell} \simeq \frac{\lambda_q}{M_h^2 - s} 
  + \frac{\lambda_{\ell} - \lambda_q}{M_H^2 - s}, 
  \qquad \qquad
  C_q \simeq \frac{\lambda_q}{M_h^2 - s}.
\end{equation}
In this limit, the $SS \to q \bar q$ channel is suppressed by keeping
$\lambda_q$ small; then the $SS \to \tau\tau$ cross section is due 
primarily to exchange of $H^0$ through the $\lambda_{\ell}$ coupling. 

For large $\tan\beta$, the upper bound on hadronic final states from
Eq.~(\ref{eq:Cqlimit}) yields the rough constraint (neglecting $s$
relative to the squared Higgs masses and neglecting $\lambda_q$
relative to $\lambda_{\ell}$ in the second term of $C_{\ell}$),
\begin{equation}
  \lambda_q \lesssim 0.14 \frac{M_h^2}{M_H^2} \lambda_{\ell}.
\end{equation}
Then, neglecting the $\lambda_q$ contributions to $C_{\ell}$, I again
obtain from Eq.~(\ref{eq:Clvalue}),
\begin{equation}
  \lambda_{\ell} \simeq \left[ \frac{M_H}{128~{\rm GeV}} \right]^2
  \left[ \frac{\sigma v_{\rm rel}}{10^{-26}~{\rm cm^3/s}} \right]^{1/2}.
  \label{eq:lambdal}
\end{equation}
In particular, taking $\sigma v_{\rm rel} = 10^{-26}$~cm$^3$/s and a
heavier second Higgs state $H^0$ of 200 (300)~GeV would require an
enhanced coupling $\lambda_{\ell} \simeq 2.4$ (5.5).  If the dark
matter annihilation cross section is near the lower end of the fitted
range of Ref.~\cite{Hooper:2010mq}, $\sigma v_{\rm rel} \simeq 3.3
\times 10^{-27}$~cm$^3$/s, this is reduced to $\lambda_{\ell} \simeq
1.4$ (3.2).  However, this low end of the annihilation cross section
range is disfavored in the model discussed here because there is no
way to boost the total annihilation cross section in the early
universe up to the range favored by the observed dark matter relic
density.  Depending on the model, such a boost can be achieved in a
number of ways: through an annihilation cross section that grows with
velocity as for $P$-wave annihilations, through co-annihilations in
the early universe with another particle carrying the dark matter
parity, or through the presence of additional annihilation final
states that do not contribute significantly to the gamma-ray flux and
thus are not included in the annihilation cross section fitted from
gamma-ray data.  None of these mechanisms apply in the model
considered here.  Alternatively, the fitted annihilation cross section
could be increased if the transition to the steeper density profile
were assumed to happen closer to the galactic center; this would in
turn favor an even lighter $H^0$ for a given $\lambda_{\ell}$.
Therefore, avoiding a large quartic scalar coupling strongly favors a
light lepton-friendly Higgs.

\section{Naturalness of the dark matter particle mass}
\label{sec:naturalness}

Reference~\cite{Hooper:2010mq} found that the gamma ray excess around
the galactic center is best fitted with a dark matter particle mass of
7.3--9.2~GeV.  The scalar potential in Eq.~(\ref{eq:VS}) leads to a mass
for the scalar $S$ in terms of the underlying model parameters of
\begin{equation}
  M_S^2 = m_3^2 + \frac{v_{\rm SM}^2}{2} 
  (\lambda_q \sin^2\beta + \lambda_{\ell} \cos^2\beta).
  \label{eq:MS}
\end{equation}
We have seen that the coupling $\lambda_{\ell}$ must be at least
$\mathcal{O}(1)$ to accommodate the fitted annihilation cross section,
while $\lambda_q$ should be at least $\sim$10 times smaller to avoid too
large an annihilation cross section to hadronic final states.

Large cancellations in Eq.~(\ref{eq:MS}) can be avoided if
$\tan\beta$ is moderately large; $\tan\beta \sim 17$ results in a
mass-squared contribution from $v_{\rm SM}^2 \lambda_{\ell}
\cos^2\beta / 2$ of about $(10~{\rm GeV})^2$.  Avoiding fine tuning
between the remaining terms requires $\lambda_q$ to be significantly
smaller than the constraint from hadronic annihilation final states; a
mass-squared contribution from $v_{\rm SM}^2 \lambda_q
\sin^2\beta / 2$ of about $(10~{\rm GeV})^2$ requires $\lambda_q \sim
1/300$.

The mass term $m_3^2$ suffers of course from the same quadratic
sensitivity to high energy scales that leads to the hierarchy problem
for the SM Higgs mass.  In this case, however, the quadratically
divergent one-loop contribution to the renormalized mass-squared comes
solely from loops of $\Phi_{\ell}$ via the coupling $\lambda_{\ell}$
(the $\Phi_q$ contribution is suppressed by the small coupling
$\lambda_q$; likewise, loops of $S$ can be suppressed by making
$\lambda_S$ small in Eq.~(\ref{eq:VS})), and is given by
\begin{equation}
  \delta M_S^2 = \frac{\lambda_{\ell}}{4 \pi^2} \Lambda^2,
\end{equation}
where $\Lambda$ is the cutoff.  While naturalness of the SM Higgs
mass-squared parameter with no more than, say, 10\% fine tuning
requires cancellation of the top quark loop quadratic divergence
around 2 TeV (for a review, see, e.g., Ref.~\cite{Schmaltz:2005ky}),
naturalness at this level for $M_S^2 \simeq (10~{\rm GeV})^2$ requires
that the $\Phi_{\ell}$ loop quadratic divergence be cancelled around
200~GeV (by, e.g., supersymmetric partners of the $\Phi_{\ell}$
states---Higgsinos around this mass scale are a common feature of
supersymmetric models).

\section{Consequences for Higgs decays}
\label{sec:invisihiggs}

The presence of a stable singlet $S$ with mass around 8~GeV and couplings 
to $h^0$ and $H^0$ has profound implications for Higgs collider 
signatures.  The partial width for $\phi \to SS$, with $\phi = h^0$ or 
$H^0$, is given by
\begin{equation}
  \Gamma(\phi \to SS) = 
  \frac{\lambda_{\phi}^2 v_{\rm SM}^2}{16 \pi M_{\phi}}
  \sqrt{1 - \frac{4 M_S^2}{M_{\phi}^2}},
\end{equation}
where the couplings for $h^0$ and $H^0$ are given by
\begin{eqnarray}
  \lambda_h &=& \lambda_q \sin\beta \cos\alpha 
  - \lambda_{\ell} \cos\beta \sin\alpha,
  \nonumber \\
  \lambda_H &=& \lambda_q \sin\beta \sin\alpha 
  + \lambda_{\ell} \cos\beta \cos\alpha.
\end{eqnarray}
I again focus on the case $\cos(\beta - \alpha) \to 0$.  The
couplings of $h^0$ and $H^0$ to scalars become,
\begin{eqnarray}
  \lambda_h &=& \lambda_q \sin^2\beta + \lambda_{\ell} \cos^2\beta,
  \nonumber \\
  \lambda_H &=& (\lambda_{\ell} - \lambda_q) \sin\beta \cos\beta.
\end{eqnarray}

I first consider the SM-like Higgs $h^0$.  The coupling $\lambda_h$
enters $C_q$ (see Eq.~(\ref{eq:Ccba}) above), so that the constraint
on hadronic annihilations leads very roughly to $\lambda_h \lesssim
0.1 \lambda_{\ell}$.  This limit is saturated for $\lambda_{\ell} \sim
1$, $\lambda_q \sim 0.1$, and $\tan\beta \sim 3$.  Neglecting the
kinematic factor $\sqrt{1 - 4 M_S^2/M_h^2}$, the invisible width of
$h^0$ is given by,
\begin{equation}
  \Gamma(h^0 \to SS) \simeq 99~{\rm MeV} 
  \left[\frac{\lambda_h}{0.1}\right]^2
  \left[ \frac{120~{\rm GeV}}{M_h} \right].
\end{equation}
When $\cos(\beta - \alpha) = 0$, the couplings of $h^0$ to SM
particles all become equal to their SM values.  The total width of the
SM Higgs, corresponding to the visible width of $h^0$, is then 3.6
(8.3, 82, 630)~MeV for $M_h = 120$ (140, 160, 180)~GeV~\cite{HDECAY}.
For $\lambda_h \sim 0.1$, then, invisible decays to $SS$ dominate the
total width of $h^0$ for masses below the $WW$ threshold.  Above the
$WW$ threshold, the invisible branching fraction falls quickly (to
about 10\% at $M_h = 180$~GeV) due to the rapid growth of the Higgs
decay widths to vector bosons.  If instead I take $\lambda_q$ and
$\tan\beta$ values as dictated by full naturalness of $M_S$, i.e.,
requiring $v_{\rm SM}^2 \lambda_h/2 \sim (10~{\rm GeV})^2$, I obtain
$\lambda_h \sim 1/300$.  This yields an invisible width for $h^0$
around 0.1~MeV, leading to an invisible branching fraction below 3\%
even for $h^0$ masses below the $WW$ threshold.  Allowing 10\% fine
tuning for the $S$ mass yields $\lambda_h \sim 1/30$ and an invisible
width for $h^0$ around 10~MeV, leading to an invisible branching
fraction of about 75\% (50\%) for $M_h = 120$ (140)~GeV.  A recent
ATLAS study~\cite{Aad:2009wy} of an invisibly-decaying Higgs produced
in vector boson fusion or associated production with a $Z$ boson finds
a 95\% confidence level exclusion reach with 30~fb$^{-1}$ down to
invisible branching fractions of about 50\% in this mass range,
assuming SM production cross sections.

Now consider the second Higgs boson $H^0$.  In order to avoid needing
too large a coupling $\lambda_{\ell}$, I will assume that $M_H$ is
below the $t \bar t$ threshold, $M_H \lesssim 350$~GeV.  Then the
partial width of $H^0$ to SM particles is dominated by $H^0 \to
\tau\tau$ for $\tan\beta \gtrsim 2$.  The partial width for $H^0 \to
\tau \tau$ is given by,
\begin{equation}
  \Gamma(H^0 \to \tau\tau) = 
  \frac{M_H m_{\tau}^2 \tan^2\beta}{8 \pi v_{\rm SM}^2} 
  \left[ 1 - \frac{4 m_{\tau}^2}{M_H^2} \right]^{3/2}
  \simeq 0.27~{\rm MeV} \, \tan^2\beta \left[\frac{M_H}{130~{\rm GeV}} \right],
\end{equation}
where in the last step I ignored the $\tau$ mass in the kinematics.
The partial width for $H^0 \to SS$ is given by
\begin{equation}
  \Gamma(H^0 \to SS) \simeq 9.2~{\rm GeV} \, \lambda_H^2 
  \left[ \frac{130~{\rm GeV}}{M_H} \right].
\end{equation}
Neglecting $\lambda_q$, taking $\tan\beta \gtrsim 3$, and applying the
fitted dark matter annihilation cross section from
Eq.~(\ref{eq:lambdal}), the coupling $\lambda_H$ is approximately given
by
\begin{equation}
  \lambda_H \simeq \cot\beta \left[\frac{M_H}{128~{\rm GeV}} \right]^2 
  \left[ \frac{\sigma v_{\rm rel}}{10^{-26}~{\rm cm^3/s}} \right]^{1/2}.
\end{equation}
Inserting this into the previous equation with $\sigma v_{\rm rel} =
10^{-26}$~cm$^3$/s, the partial width for $H^0 \to SS$ becomes,
\begin{equation}
  \Gamma(H^0 \to SS) \simeq 9.8~{\rm GeV} \cot^2\beta 
  \left[ \frac{M_H}{130~{\rm GeV}} \right]^3.
\end{equation}

Sufficient suppression of hadronic final states in $SS$ annihilation
requires $\tan\beta \gtrsim 3$, while full naturalness of $M_S$ favors
$\tan\beta \sim 17$.  Taking $M_H = 130$~GeV and $\tan\beta = 3$
yields an overwhelmingly invisibly decaying $H^0$, with only about 2
per mille visible decays to $\tau\tau$.  Taking $\tan\beta = 17$
yields an invisible branching fraction for $H^0$ of about 30\% for
this mass, with the remainder of decays to $\tau\tau$.  The ratio of
invisible to visible widths grows with the $H^0$ mass like $M_H^2$,
yielding an invisible branching fraction of about 50\% at $M_H =
200$~GeV; this growth is due to the larger coupling $\lambda_{\ell}$
required to obtain the fitted dark matter annihilation cross section
at higher $H^0$ masses.  Allowing 10\% fine tuning for the $S$ mass
yields $\tan\beta \sim 5.5$ for $\lambda_{\ell} \sim 1$, leading to a
visible width (to $\tau\tau$) of a 130~GeV $H^0$ of about 2.5\%, with
the remainder of decays invisible.

Discovery of an invisibly-decaying $H^0$ at the LHC is a major
challenge.  When $\cos(\beta - \alpha) \to 0$, $H^0$ is not produced
in vector boson fusion or in association with a $W$ or $Z$ boson.  Its
cross sections in gluon fusion and $t \bar t H^0$ associated
production are suppressed by $\cot^2\beta$.  Instead, its primary
production modes are through $s$-channel $Z$ ($W$) exchange in
association with a CP-odd (charged) Higgs, which decays predominantly
to $\tau\tau$ ($\tau\nu$).  Discovery and characterization of such a
particle may have to wait for a high-energy $e^+e^-$ collider.  Because
electroweak precision constraints tend to disfavor large mass splittings
among the members of a second Higgs doublet, the International Linear
Collider operating with a center-of-mass energy around 500~GeV should
be able to study this model in detail (for a review of International
Linear Collider capabilities for the second Higgs doublet of the minimal
supersymmetric standard model, see, e.g., Ref.~\cite{Djouadi:2007ik}).

\section{Direct detection cross section}
\label{sec:direct}

Scalar dark matter particles $S$ in the galactic halo can scatter off
nuclei via exchange of $h^0$ and $H^0$, yielding direct recoil signals
for dark matter.  In this process, $h^0$ and $H^0$ must couple to the
quarks or gluons in the nucleus; the scattering cross section is thus
suppressed by the same physics that yields the suppression of
annihilations to hadronic final states.  Following, e.g.,
Ref.~\cite{Barger:2010yn}, the spin-independent cross section for $S$
scattering off the proton is
\begin{equation}
  \sigma_{\rm SI} = \frac{m_p^4}{2 \pi (m_p + M_S)^2} C_q^2
  \left[ f_{pu} + f_{pd} + f_{ps} + \frac{2}{27}(3 f_G) \right]^2,
\end{equation}
where $m_p$ is the proton mass, $C_q$ is the coupling and propagator
factor given in Eq.~(\ref{eq:Cs}) evaluated at $s = 0$, and the nuclear
formfactors in the square brackets are given by~\cite{Ellis:2000ds},
\begin{equation}
  f_{pu} = 0.02, \qquad f_{pd} = 0.026, \qquad f_{ps} = 0.118, \qquad
  f_G = 0.836,
\end{equation}
with an uncertainty of about 30\%.

Neglecting $(2 M_S)^2$ compared to $M_h^2$, $M_H^2$ in the coupling
and propagator factors, this can be expressed in terms of the
parameter region favored by the annihilation signal according to,
\begin{equation}
  \sigma_{\rm SI} \simeq 4.0 \times 10^{-42}~{\rm cm}^2 
  \left[ \frac{9~{\rm GeV}}{m_p + M_S} \right]^2 
  \left[ \frac{\sigma v_{\rm rel}}{10^{-26}~{\rm cm^3/s}} \right]
  \left[ \frac{C_q^2/C_{\ell}^2}{0.015} \right],
\end{equation}
where the term in the last set of square brackets is at most equal to
1 due to the upper bound of $\sim 20\%$ on the fraction of dark matter
annihilations to hadrons.  We thus obtain an upper bound on the direct
detection cross section roughly an order of magnitude below the cross
section required for the (controversial) dark matter interpretation of
the CoGeNT data~\cite{Aalseth:2010vx}.  This cross section is likewise
about a factor of 5 below the current upper limit from the XENON100
experiment~\cite{Aprile:2010um} for $M_S = 8$~GeV.

\section{Conclusions}
\label{sec:conclusions}

Hooper and Goodenough~\cite{Hooper:2010mq} identified an excess
component of the gamma ray flux near the galactic center as measured
by the Fermi Gamma Ray Space Telescope and showed that it is well
fitted by annihilations of $\sim$8~GeV dark matter particles into
$\tau\tau$ final states.  I showed in this paper that this scenario
can be realized in the lepton-specific two-Higgs-doublet model
extended by a stable gauge-singlet scalar dark matter particle.

Obtaining the right annihilation cross section requires
$M/\sqrt{\lambda_{\ell}} \sim 130$~GeV, where $M$ is the mass of the
CP-even neutral Higgs state that lives mostly in the
``lepton-friendly'' Higgs doublet $\Phi_{\ell}$ and $\lambda_{\ell}$
is the four-point coupling of $\Phi_{\ell}$ to dark matter particle
pairs.  Avoiding too large a four-point coupling thus forces the
lepton-friendly Higgs to be light.  Suppressing annihilations to
quarks requires both the four-point coupling of the ``quark-friendly''
doublet to dark matter and the mixing between the two doublets to be
small.  Complete naturalness of the rather low dark matter particle
mass can be achieved if the vacuum expectation value of $\Phi_{\ell}$
is around 14~GeV (i.e., $\tan\beta \sim 17$) and the four-point
coupling of the quark-friendly doublet to dark matter is around 1/300
or less---in itself a rather fine-tuned situation.  Allowing 10\% 
fine tuning for the $S$ mass leads to the more comfortable values of
$\tan\beta \sim 5.5$ and $\lambda_q \lesssim 1/30$.

The most significant consequences of this scenario for collider
physics are as follows.  First, both of the CP-even neutral Higgs
states should be in the 100--200~GeV mass range in order to avoid too
large a coupling $\lambda_{\ell}$.  Second, in part of the allowed
parameter space, both of the CP-even neutral Higgs states will decay
predominantly to dark matter particles; requiring no more than 10\%
fine tuning for the dark matter particle mass, the Standard Model-like
Higgs acquires an invisible width comparable to its visible width,
while the lepton-friendly Higgs decays mostly invisibly with a
branching fraction to $\tau\tau$ of a few percent.  Because of the
suppression of their single-production couplings to quarks and gauge
bosons, discovery of the four Higgs states of the lepton-friendly
doublet in this scenario is a major challenge at the LHC, but should
be straightforward at a $\sim$500~GeV International Linear $e^+e^-$
Collider.

\begin{acknowledgments}
I thank Thomas Gregoire for helpful discussions.
This work was supported by the Natural Sciences and Engineering
Research Council of Canada.
\end{acknowledgments}


\end{document}